\documentclass[PRD,aps,twocolumn,showpacs,preprintnumbers,amsmath,amssymb,superscriptaddress,nofootinbib]{revtex4-1}
\usepackage[dvips]{graphicx}
\usepackage{epsf}
\usepackage{amsmath}
\usepackage{amssymb}
\usepackage{amsfonts}
\usepackage{times}
\usepackage[usenames]{color}
\usepackage[normalem]{ulem}
\usepackage[T1]{fontenc}
\usepackage[dvipsnames]{xcolor}

\usepackage{graphicx}
\usepackage{dcolumn}
\usepackage{bm}
\usepackage{color}

\begin{document}

\title{Subhalo abundance matching in $f(R)$ gravity}

\author{Jian-hua He}
\email[Email address: ]{jianhua.he@durham.ac.uk}
\affiliation{Institute for Computational Cosmology, Department of Physics, Durham University, Durham DH1 3LE, UK}

\author{Baojiu Li}
\affiliation{Institute for Computational Cosmology, Department of Physics, Durham University, Durham DH1 3LE, UK}

\author{Carlton M. Baugh}
\affiliation{Institute for Computational Cosmology, Department of Physics, Durham University, Durham DH1 3LE, UK}

\begin{abstract}
Using the liminality N-body simulations of Shi et. al., we present the first predictions for galaxy clustering in $f(R)$ gravity using subhalo abundance matching. We find that, for a given galaxy density, even for an $f(R)$ model with $f_{R0}=-10^{-6}$, for which the cold dark matter clustering is very similar to $\Lambda$CDM, the predicted clustering of galaxies in the $f(R)$ model is very different from $\Lambda$CDM. The deviation can be as large as $40\%$ for samples with mean densities close to that of $L_*$ galaxies. This large deviation is testable given the accuracy that future large-scale galaxy surveys aim to achieve. Our result demonstrates that galaxy surveys can provide a stringent test of General Relativity on cosmological scales, which is comparable to the tests from local astrophysical observations.
\end{abstract}

\maketitle
{\bf Introduction.}~With the advent of ever larger galaxy redshift surveys,  there has been a steady
improvement in the accuracy of measurements of galaxy clustering. Upcoming large galaxy surveys such as the Dark Energy Spectroscopic Instrument (DESI) survey~\cite{DESI} and the Euclid mission~\cite{Euclid} aim to achieve percent level accuracy on the measurement of galaxy clustering. The measurements will not only provide an unprecedented constraint on the cosmological parameters in $\Lambda$CDM but, as we show,  can also produce a stringent test of General Relativity on cosmological scales.

Modified gravity models change the merger histories and distributions of cold dark matter subhalos, leading to different clustering of subhalos. Matching galaxies to subhalos is, therefore, a good way to test modified gravity models, which can avoid having to modelling the complicated baryonic physics in different gravity models.  In its original form, the subhalo abundance matching (SHAM) method assumes that there is a one-to-one relationship between a property of a subhalo and an observable property of a galaxy. The galaxy property is usually taken to be the stellar mass. However, unlike the more strongly gravitationally bound stellar component of galaxies, the dark matter of a subhalo can suffer from tidal stripping. A ``satellite'' subhalo can lose a substantial amount of mass depending on how close its orbit approaches the central part of the halo. The subhalo number density profile is therefore much shallower than the number density profile of galaxies in hydrodynamical simulations as well as the number density profile of galaxies in observations (see Fig.2 in Ref.~\cite{galaxysim}). Galaxy properties (here stellar mass) are expected to be closely connected to host subhalo properties at some epoch before this disruption. Reference~\cite{acc} suggests connecting a galaxy's stellar mass to a subhalo's  maximum circular velocity $v_{\rm max}={\rm Max}[ \sqrt{GM(<r)/r}]$ at the epoch of accretion (hereafter $v_{\rm acc}$), where $M$ is the mass of the subhalo. In a pure cold dark matter simulation, using $v_{\rm acc}$ can result in more subhalos being selected from the central region of a host halo and yields a much steeper subhalo number density profile, which in turn leads to a better fit to the observed galaxy clustering on small scales. An alternative method is to find the peak value of the maximum circular velocity (i.e. the maximum value of $v_{\rm max}$) over a subhalo's merger history~\cite{vpeak,vpeakfit,qiguo} (hereafter $v_{\rm peak}$). Since the maximum circular velocity $v_{\rm max}$ is closely related to the self-gravity of the subhalo, a subhalo at the epoch of $v_{\rm peak}$ has the strongest binding force and, thereby, is most stable against tidal stripping. The subhalo properties therefore are expected to be tightly correlated with the galaxy stellar mass at this epoch.  This point is partially confirmed by the state-of-the-art hydrodynamic simulation EAGLE~\cite{EAGLE,EAGLEsham}. The correlation between $v_{\rm peak}$ of a subhalo and galaxy stellar mass links theory to observation. If we assume galaxies reside in subhalos, through a monotonic relation, subhalos selected by $v_{\rm peak}$ in a simulation should correspond to galaxies selected by stellar mass in a galaxy survey. The predicted clustering in a simulation can, therefore, be compared to the observed clustering directly. SHAM, thus, provides a straightforward way to test different gravity models.

Here, we investigate, for the first time, galaxy clustering in modified gravity models using the subhalo abundance matching method. For illustrative purposes we focus on $f(R)$ gravity which is one of the most popular modified gravity models (see Ref.~\cite{frreview} for review). $f(R)$ gravity introduces an extra scalar degree of freedom, which mediates a fifth force that changes the motion of massive particles. However, it also incorporates a screening mechanism which can suppress this fifth force in high-density environments \cite{Khoury}, therefore mimicking GR in environments such as our solar system and the early Universe. These effects can only be addressed using $N$-body simulations, which requires an explicit functional form for $f(R)$. We choose the Hu-Sawicki model~\cite{HuS} with the index $n=1$~\cite{HuS}. In order to illustrate the robustness of the SHAM method, we choose the free parameter in the $f(R)$ model as $f_{R0}=-10^{-6}$, for which the model closely resembles $\Lambda$CDM. The relative difference in the non-linear cold dark matter power spectrum between this $f(R)$ model and $\Lambda$CDM is less than $5\%$ up to $k\sim 10 h{\rm Mpc}^{-1}$ (see e.g. Ref.~\cite{ECOSMOG}). Hence, this model can hardly be distinguished from $\Lambda$CDM using current cosmological probes such as the number counts of clusters~\cite{clusterconstraints} or weak lensing~\cite{lensing}. The model has only been tested with local astrophysical observations (e.g.\cite{astro_constraint}). However, we shall show that this model still leaves a significant signature in galaxy clustering and can be robustly tested on cosmological scales. 

{\bf N-body simulations}~We use the liminality simulations presented in Ref.~\cite{Shidifu} for the Hu-Sawicki model~\cite{HuS}. The simulations were performed using the {\sc ecosmog} code~\cite{ECOSMOG} which is based on the $N$-body code {\sc ramses}~\cite{RAMSES}. The box size is $L_{\rm box}=64 h^{-1}{\rm Mpc}$.  The cosmological parameters are $\Omega_b^0=0.046, \Omega_c^0=0.235, \Omega_d^0=0.719, h=0.697, n_s=0.971$, and $\sigma_8=0.820$. The number of particles is $N=512^3$ and the mass resolution is $m_p=1.52\times 10^8h^{-1}M_{\odot}$ which is the highest resolution cosmological simulation to date of the $f(R)$ model considered here. The simulation has $122$ snapshots between $z=49$ to $z=0$. In addition to the $f(R)$ simulation, we use a $\Lambda$CDM simulation with the same box size, resolution, cosmological parameters, initial conditions and number of snapshots as for comparison.

\begin{figure*}
\includegraphics[width=6.12in,height=2.35in]{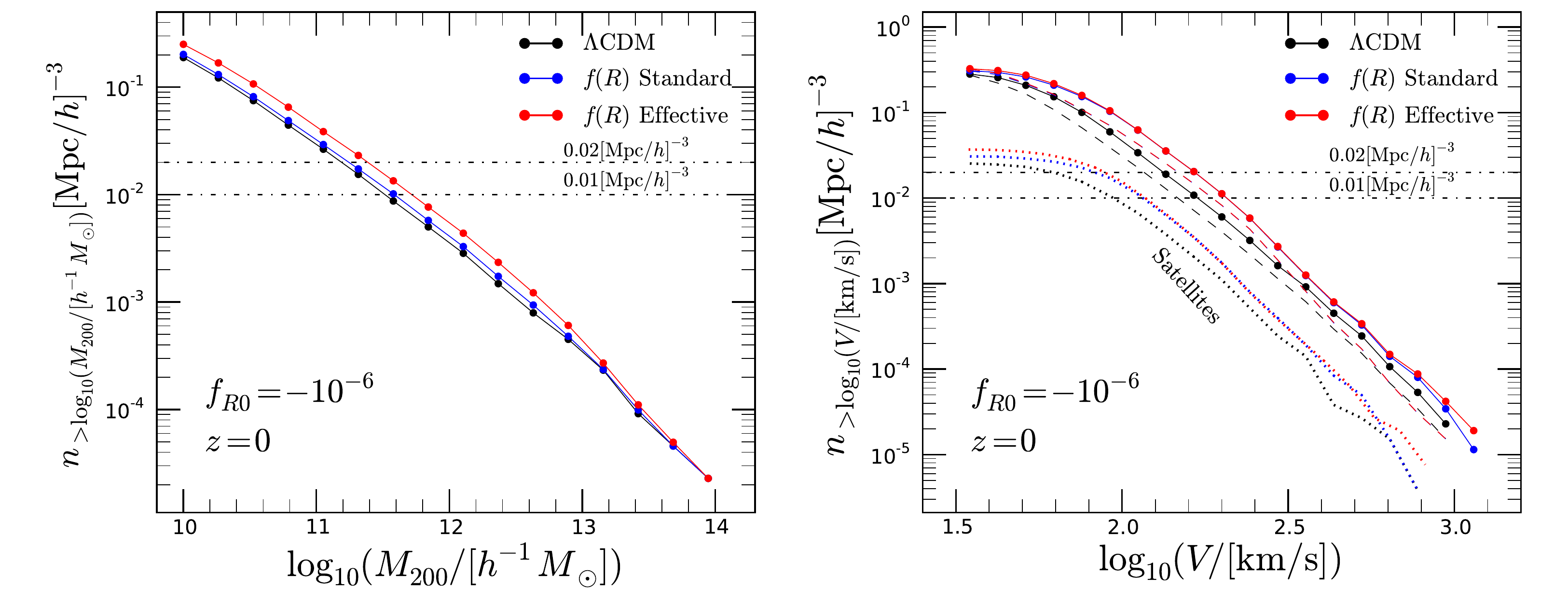}
\caption{The global cumulative subhalo abundance as a function of halo mass $M_{200}$ (left), and the current maximum circular velocity $v_{\rm max}$ and $v_{\rm peak}$ (right). Here the subhalo catalogs include both satellite and main subhalos.  In $\Lambda$CDM, the cumulative subhalo number counts as a function of mass are well approximated by a power law. However, the mass function in the $f(R)$ model is more complicated. At the high mass end, $M>10^{13} h^{-1}M_{\odot}$, due to the efficient screening, the cumulative subhalo number counts in the $f(R)$ model are very close to those in $\Lambda$CDM. However, at the low mass end, the abundance of subhalos in the $f(R)$ model is higher than in $\Lambda$CDM due to the enhanced gravity in unscreened halos. The enhancement is more significant in the effective halo catalog of the $f(R)$ model since the effective mass is used in this case. For the subhalo abundance plotted in terms of $v_{\rm max}$ and $v_{\rm peak}$ (right), compared with $v_{\rm max}$ (dashed lines), using $v_{\rm peak}$ (solid lines) enhances the overall abundance of subhalos in the $f(R)$ model even for the most massive ones. In the right panel, the dotted lines show the abundance of satellite subhalos (excluding main subhalos) selected using $v_{\rm peak}$ for $\Lambda$CDM (black), the $f(R)$ standard halo catalog (blue) and the $f(R)$ effective halo catalog (red).  The abundance of satellite subhalos is relatively complete for the full subhalo samples investigated with mean number densities $<n_g>=0.01[{\rm Mpc}/h]^{-3}$ and $<n_g>=0.02[{\rm Mpc}/h]^{-3}$. } \label{one}\label{velocityfunction}
\end{figure*}

In order to perform subhalo abundance matching, we need two crucial pieces of information: the maximum circular velocity $v_{\rm max}$ and the merger history of subhalos. Unlike the case of $\Lambda$CDM, $v_{\rm max}$ in $f(R)$ gravity is not directly related to the true cold dark matter mass of a subhalo but to an effective mass which is defined through the modified Poisson equation ~\cite{EFFhalos}
\begin{equation}
\nabla \phi=4\pi G a^2 \delta \rho_{\rm eff}\quad,
\end{equation}
where $G$ is Newton's constant. The effective energy density $\delta\rho_{\rm eff}$, by definition, incorporates all the effects of modified gravity. The circular velocity is then given by $$v_{\rm cir}^2(r)=\frac{GM_{\rm eff}(<r)}{r}\quad,$$ where $M_{\rm eff}$ is the effective mass enclosed within a radius of $r$ for a dark matter halo. We therefore need to build effective halo catalogs~\cite{EFFhalos} from the $f(R)$ simulation. The details are presented in Ref~\cite{EFFhalos,freffpk} in which the halos are identified using a modified version of the {\sc Amiga} Halo Finder ({\sc Ahf})~\cite{AHF}. We build the halo merger tree using the {\sc mergertree} code which is part of the {\sc Ahf} package.  For comparison we also consider the standard halo catalog for the $f(R)$ simulation. The standard halo catalog is simply built from the density field of cold dark matter. However, we calculate $v_{\rm max}$ taking into account the modification of gravity. Thus, in both the effective and standard halo catalogs, $v_{\rm max}$ is physically defined. 

{\bf Subhalo abundance}~We show in Fig.~\ref{one} the cumulative abundance of subhalos as a function of halo mass $M_{200}$ (left panel), and the current maximum circular velocity $v_{\rm max}$ and $v_{\rm peak}$ (right panel). The halo catalog used here includes both satellite and main subhalos. Note that the $f(R)$ standard halo catalog uses the true cold dark matter mass while the $f(R)$ effective halo catalog uses the effective mass for subhalos. The left panel of Fig.~\ref{one} shows that the cumulative mass function in $\Lambda$CDM is well approximated by a power law. However, the mass function in the $f(R)$ model has a more complicated shape. At the high mass end,  $M>10^{13} h^{-1}M_{\odot}$,  the subhalo number counts in the $f(R)$ model are very close to those in $\Lambda$CDM while at the low mass end, $M<10^{13} h^{-1}M_{\odot}$, the abundance in the $f(R)$ model is higher than in $\Lambda$CDM. This is due to the screening mechanism in the $f(R)$ model. A massive halo in the $f(R)$ model is usually screened. There are no significant differences between a screened $f(R)$ halo and  a $\Lambda$CDM halo of the same mass. However, a low mass halo in the $f(R)$ model is usually unscreened and experiences enhanced gravity. The enhanced gravity can speed up the halo assembly and therefore increases the abundance of halos of a given mass.  Enhanced gravity also leads to a greater effective mass, which is the dominant effect. This is why the enhancement is more significant in the effective halo catalog of the $f(R)$ model as shown in Fig.~\ref{one}.
\begin{figure*}
\includegraphics[width=6.12in,height=2.35in]{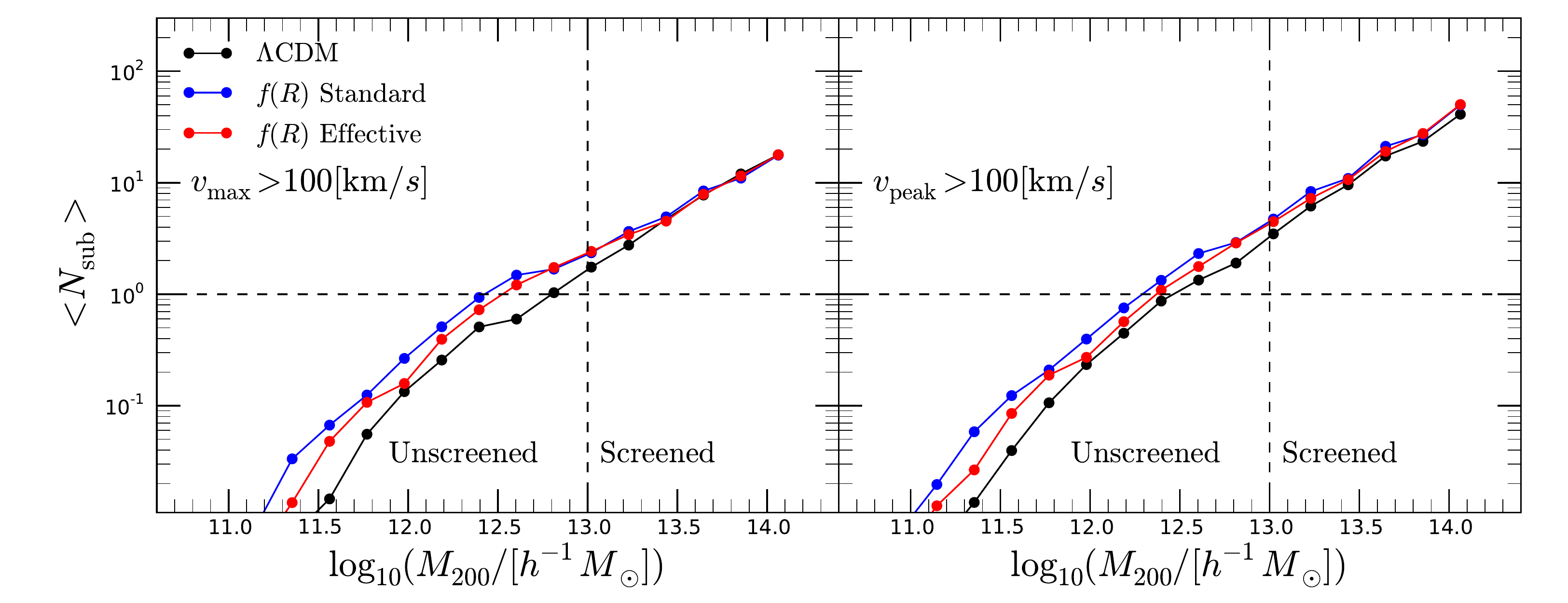}
\caption{The mean number of selected satellite subhalos (excluding main subhalos) per host halo as a function of the mass of their host halo for $\Lambda$CDM (black points), the $f(R)$ standard halo catalog (blue points) and the $f(R)$ effective halo catalog (red points), respectively. Left: subhalos are selected by their current maximum circular velocity $v_{\rm max}>100 {\rm km/s}$.  
At the high mass end, $M>10^{13} h^{-1}M_{\odot}$, due to the screening mechanism in the $f(R)$ model, the mean satellite subhalo occupations of the $f(R)$ model and $\Lambda$CDM are very similar to one another and the mean number of selected satellite subhalos per host halo is proportional to the mass of their host halos $<N_{\rm sub}>\propto M$. However, at the low mass end, using $v_{\rm max}$ tends to select more satellite subhalos per host halo in the $f(R)$ model than in $\Lambda$CDM since unscreened halos in the $f(R)$ model experience enhanced gravity and therefore a boosted value of $v_{\rm max}$.
Right: similar to the left panel but subhalos are selected by $v_{\rm peak}$. In contrast to $v_{\rm max}$, using $v_{\rm peak}$ enhances the overall selection of subhalos in the $f(R)$ model even for the most massive ones. }\label{subHODs}
\end{figure*} 

In the right panel of Fig.~\ref{one}, we show the abundance of subhalos measured in terms of $v_{\rm max}$ and $v_{\rm peak}$. In contrast to halo masses, the abundances of subhalos measured by $v_{\rm max}$ and $v_{\rm peak}$ in the standard catalog (blue curves) and the effective catalog (red curves) are very close to one another. This is because the circular velocity of subhalos in the standard catalog is calculated taking into account the modification of gravity. Compared with $v_{\rm max}$, using $v_{\rm peak}$ yields a higher abundance of subhalos for both the $f(R)$ model and $\Lambda$CDM. However, unlike $v_{\rm max}$,  using $v_{\rm peak}$ enhances the abundance of massive  screened subhalos in the $f(R)$ model as well. This is due to the selection effect of $v_{\rm peak}$ and the fact that before a satellite subhalo merges into a screened massive host halo, the satellite subhalo can be a  distinct low mass unscreened main halo. In order to address this point, in Fig.~\ref{subHODs}, we plot the mean number density of satellite subhalos (excluding main subhalos) as a function of the mass of their host halo. In the left panel of Fig.~\ref{subHODs}, subhalos are selected by their current maximum circular velocity so that $v_{\rm max}>100 {\rm km/s}$.  It can be seen that, at the high mass end, $M>10^{13} h^{-1}M_{\odot}$, due to the screening mechanism in the $f(R)$ model, the mean satellite subhalo occupations for the $f(R)$ model and $\Lambda$CDM are very similar to one another and the mean number of selected satellite subhalos per host halo is proportional to the mass of the host halo $<N_{\rm sub}>\propto M$, consistent with Ref.~\cite{subhaloHOD}. However, at the low mass end, $M<10^{13} h^{-1}M_{\odot}$, using $v_{\rm max}$ to select subhalos tends to recover more satellite subhalos per host halo in the $f(R)$ model than in $\Lambda$CDM since the unscreened halos in the $f(R)$ model experience enhanced gravity and therefore a boosted value of $v_{\rm max}$. However, as shown in the right panel of Fig.~\ref{subHODs}, using $v_{\rm peak}$ enhances the overall selection of subhalos in the $f(R)$ model not only for the unscreened halos but also for the massive screened ones.
\begin{figure*} 
\includegraphics[width=6.12in,height=2.52in]{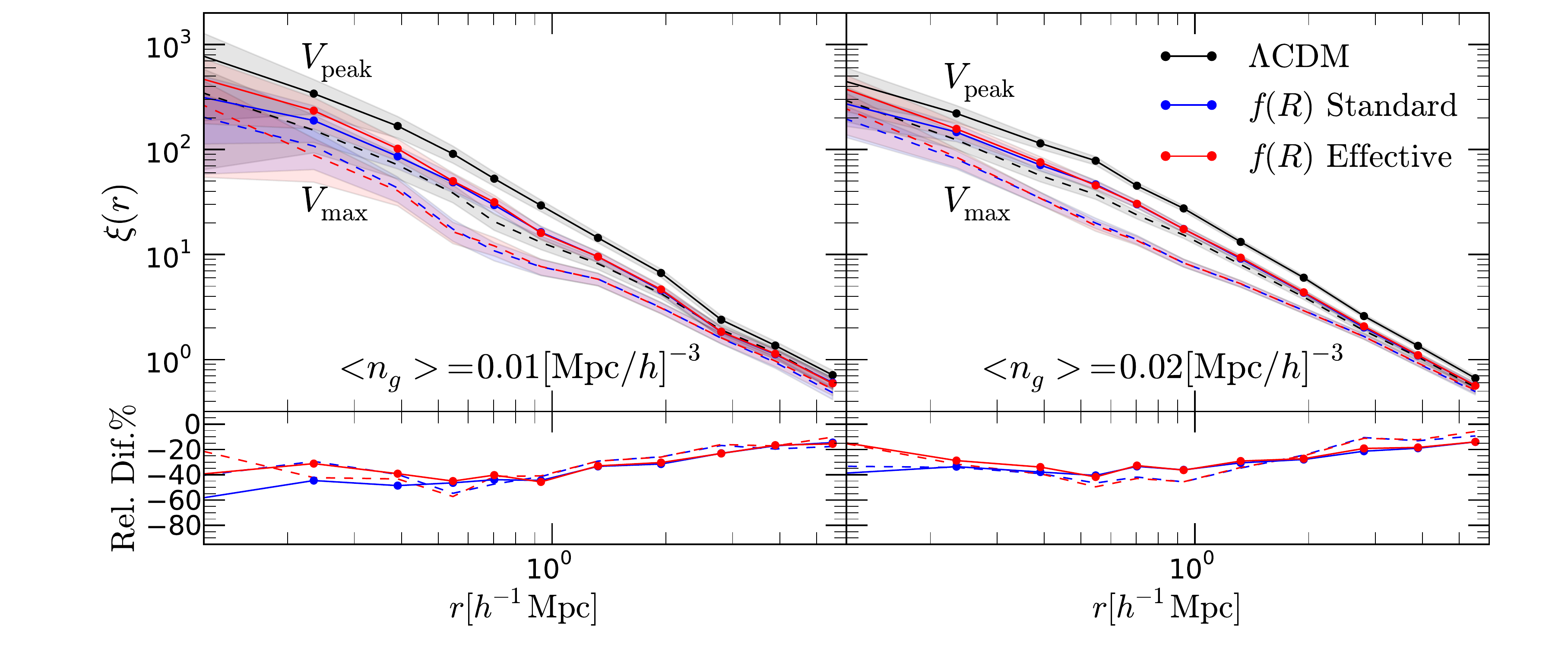}
\caption{The predicted three-dimensional two-point galaxy correlation functions from the SHAM model (upper panels). The shaded regions represent the $1\sigma$ Poisson errors. The lower panels show the fractional differences between the $f(R)$ model and $\Lambda$CDM. The left panels show the results for a galaxy density $<n_g>=0.01[{\rm Mpc}/h]^{-3}$ and the right panels are for $<n_g>=0.02[{\rm Mpc}/h]^{-3}$ . For comparison, the dashed lines shows the results obtained using the current maximum circular velocity $v_{\rm max}$.  }\label{galaxyCF}
\end{figure*}
 
{\bf SHAM clustering predictions}
SHAM assumes that there is a monotonic relation between $v_{\rm peak}$ of a subhalo and galaxy stellar mass. We therefore select subhalos in our halo catalogs by ranking them in terms of $v_{\rm peak}$. By further assuming that the selected subhalos have a one-to-one correspondence to galaxies, in Fig.~\ref{galaxyCF} we show the predicted three-dimensional galaxy two-point correlation functions for two different representative galaxy densities (upper panels) as well as the fractional differences between the $f(R)$ model and $\Lambda$CDM $(\xi_{f(R)}/\xi_{\Lambda \rm{CDM}}-1)\times 100\%$ (lower panels). Our measurements of $\xi(r)$ use the {\sc CUTE} code~\cite{cutecode}. For comparison, we also present the galaxy clustering predicted using the current maximum circular velocity $v_{\rm max}$ to rank subhalos (dashed lines). The shaded regions in Fig.~\ref{galaxyCF} represent $1\sigma$ Poisson errors. Note that due to the limited box size of our simulations, we can only measure $\xi(r)$ on scales $r<0.1L_{\rm box} \approx 6.4 h^{-1}{\rm Mpc}$. 

The upper panels of Fig.~\ref{galaxyCF} show that overall the predicted galaxy clustering in the $f(R)$ model using both $v_{\rm peak}$ and $v_{\rm max}$ is significantly weaker than in $\Lambda$CDM.  
This is because in $f(R)$ gravity subhalos in smaller unscreened main halos are more likely to have higher $v_{\rm peak}$ and $v_{\rm max}$ than subhalos of equivalent masses in large screened main halos due to the enhanced gravity in the former case. As a result, the overall effect of using $v_{\rm peak}$ or $v_{\rm max}$ is to preferentially select more subhalos from less massive unscreened halos in the $f(R)$ model compared to $\Lambda$CDM.  Since subhalos in less massive host halos are less clustered, the clustering in the $f(R)$ model is expected to be weaker than in $\Lambda$CDM. 

From Fig.~\ref{galaxyCF}, it is also interesting to note that even for the $f(R)$ model with $f_{R0}=-10^{-6}$, for which the clustering of the cold dark matter is essentially indistinguishable from $\Lambda$CDM (e.g. see Ref.~\cite{ECOSMOG}), the predicted galaxy clustering in the $f(R)$ model shows sizeable reductions from $\Lambda$CDM. The maximum reduction is about $40\%$ for both the $<n_g>=0.01[{\rm Mpc}/h]^{-3}$ and $<n_g>=0.02[{\rm Mpc}/h]^{-3}$ samples. Moreover, the relative deviations are significant given the statistical errors as shown in Fig.~\ref{galaxyCF}.

In addition to the differences between the $f(R)$ model and $\Lambda$CDM, the predicted galaxy clustering in the effective halo catalog and the standard halo catalog also show differences on small scales. This is expected since the two halo catalogs are essentially different. The differences are due to the different definitions of halo centres as well as the different halo abundances (see, Fig.~\ref{velocityfunction}). However, on large scales, the two catalogs yield convergent results.  

{\bf Summary} Using the liminality simulations presented in Ref.~\cite{Shidifu}, we have studied the SHAM predictions for galaxy clustering in $f(R)$ gravity. We find that, for a given galaxy density, even for the $f(R)$ model with $f_{R0}=-10^{-6}$, for which the clustering of cold dark matter is very similar to $\Lambda$CDM, the predicted galaxy clustering in the $f(R)$ model is much weaker than in $\Lambda$CDM. The deviation can be as large as $40\%$ for samples with $<n_g>=0.01[{\rm Mpc}/h]^{-3}$ which correspond to brighter galaxy samples as well as samples with $<n_g>=0.02[{\rm Mpc}/h]^{-3}$ which correspond to slightly fainter galaxy samples. Moreover, the relative deviations are robust against statistical errors and the results obtained using $v_{\rm peak}$ and $v_{\rm max}$ in both the effective and standard halo catalogs are convergent on scales $r> 0.6 h^{-1}{\rm Mpc}$.

In modern applications of SHAM, a scatter is usually added between $v_{\rm peak}$ of a subhalo and galaxy stellar mass. However, the scatter indeed has a limited effect on our results. First, the scatter is constrained to some extent by observations such as the baryonic Tully-Fisher relation (or its equivalent for early-type galaxies) (e.g. Ref.~\cite{LV}), and is usually taken as a fixed value (e.g Ref.~\cite{boss}). Second, the scatter only affects the selection of subhalos around a mass cut. Therefore, it only affects the clustering of subhalos with very low number densities. As shown in Ref.~\cite{vpeakfit}, for high density samples, such as those investigated in this work, the impact of scatter within the range allowed by observations is negligible.
 
Another factor that might affect our results is baryonic physics. However, based on the state-of-the-art hydrodynamical simulation EAGLE~\cite{EAGLE}, which can reasonably reproduce the observed galaxy properties, Ref.~\cite{EAGLEsham} found that the agreement between the predicted galaxy clustering using SHAM and the simulated galaxy clustering is better than $30\%$ on small scales $r<1\,h^{-1}{\rm Mpc}$ and better than $10\%$ on scales larger than $r>1.3\,h^{-1}{\rm Mpc}$. The deviation shown here between the $f(R)$ model and $\Lambda$CDM is much larger than this uncertainty and therefore these models should be distinguishable. Our results therefore indicate that galaxy surveys, on cosmological scales, have the potential to constrain modified gravity models at a similar level to the local astrophysical tests (e.g.~\cite{astro_constraint}).

Moreover, the SHAM predictions can be practically tested against current and upcoming observations. SHAM predictions can be directly compared with the galaxy clustering measured from a volume-limited sample that is complete in stellar mass. The sample can be constructed from current available data sets such as the main galaxy sample of the Sloan Digital Sky Survey (SDSS)~\cite{DR7} in which the number densities of the faint galaxies can be as high as $<n_g>=0.03[{\rm Mpc}/h]^{-3}$ covering
the densities investigated in this work. A detailed analysis using the SDSS data will be presented in a separate paper. The survey
area of the bright galaxy samples (BGS) from the upcoming DESI survey~\cite{DESI} is twice as large
as that of the SDSS main galaxy sample, and can provide better
statistics for testing not only the projected galaxy clustering but also the redshift-space galaxy clustering.

{\bf Acknowledgments} We thank Difu Shi for sharing the simulation data. J.H.H. acknowledges support of the Durham International Junior Research Fellowship RF040426. B.L. acknowledges support by the UK STFC Consolidated Grant ST/L00075X/1 and RF040335. CMB acknowledges a research fellowship from the Leverhulme Trust. This work has used the DiRAC Data Centric system at Durham University, operated by the Institute for Computational Cosmology on behalf of the STFC DiRAC HPC Facility (www.dirac.ac.uk). This equipment was funded by BIS National E-infrastructure capital grant ST/K00042X/1, STFC capital grant ST/H008519/1, STFC DiRAC Operations grant ST/K003267/1 and Durham University. DiRAC is part of the National E-Infrastructure. For
access to the halo catalogs, please contact J.H.H.

\end{document}